\documentclass[superscriptaddress,preprintnumbers,showpacs]{revtex4}
\usepackage{amssymb,graphicx,dcolumn,bm}

\begin{document}

\newcommand*{\PKU}{School of Physics and State Key Laboratory of Nuclear Physics and
Technology, \\Peking University, Beijing 100871}\affiliation{\PKU}

\title{Parton distribution functions and nuclear EMC effect \\in a
statistical model\footnote{Talk given at the 5th International
Conference On Quarks and Nuclear Physics (QNP09), Sep 2009, Beijing
China, published in Chin.Phys.C34:1417-1420,2010.}}

\author{Lijing Shao}\affiliation{\PKU}
\author{Yunhua Zhang}\affiliation{\PKU}
\author{Bo-Qiang Ma}\email{mabq@phy.pku.edu.cn}\affiliation{\PKU}

\begin{abstract}
A new and simple statistical approach is performed to calculate the
parton distribution functions (PDFs) of the nucleon in terms of
light-front kinematic variables. Analytic expressions of
$x$-dependent PDFs are obtained in the whole $x$ region. And
thereafter, we treat the temperature $T$ as a parameter of the
atomic number $A$ to explain the nuclear EMC effect in the region $x
\in [0.2, 0.7]$. We give the predictions of PDF ratios, and they are
very different from those by other models, thus experiments aiming
at measuring PDF ratios are suggested to provide a discrimination of
different models. 
\end{abstract}


\maketitle

\section{Introduction}

The nucleon structure functions, in terms of the parton distribution
functions (PDFs), are badly desired in hadronic study. However, due
to the complicated non-perturbative effect, we still have difficulty
to calculate them absolutely from the first principal theory of
quantum chromodynamics (QCD).

Various models according to the spirit of QCD have been brought
forward, therein statistical ones, providing intuitive appeal and
physical simplicity, have made amazing success~\cite{lab1, lab2,
lab3, lab4, lab5, lab6, lab7, lab8, lab9, lab10, lab11, lab12,
lab13, lab14, lab15, lab16, lab17, lab18, lab19, lab20, lab21,
lab22, lab23, lab24, lab25, lab26, lab27, lab28, lab29}. Actually,
as can be speculated, with partons bound in the wee volume of the
nucleon, not only the dynamic, but also statistical properties, for
example, the Pauli exclusion principle, should have important effect
on the PDFs.

In order to avoid tough problems risen in the infinite-momentum
frame (IMF)~\cite{lab30,lab31,lab32}, we start with instant-form
statistical expressions in the nucleon rest frame, then perform
transformation in terms of light-front kinematic variables. The
analytic expressions of the PDFs we get are somehow different from
those attained in other statistical models implemented in the
IMF~\cite{lab4, lab5, lab6, lab7, lab8}, and ours perform better
with non-vanishing PDFs when $x \rightarrow 0$.

On the other hand, the nucleons in a nucleus were initially thought
to be highly insensitive to their surroundings, and the only nuclear
effect in deep inelastic scattering (DIS) was believed to be Fermi
motion at large $x$. However, in 1982, it was discovered that
nucleons inside a nucleus have a remarkably different momentum
configuration as expected, which was named nuclear EMC
effect~\cite{lab33, lab34, lab35, lab36}. In order to account for
the EMC effect, there have been many efforts and insights
implemented in various models, e.g., the cluster model~\cite{lab37,
lab38, lab39, lab40}, the pion excess model~\cite{lab38, lab41,
lab42, lab43}, the $x$-rescaling model~\cite{lab44, lab45}, the
$Q^2$-rescaling model~\cite{lab46, lab47, lab48}, and the nucleon
swelling model~\cite{lab49}. The statistical idea is also applied to
the EMC effect~\cite{lab50, lab51, lab52}. However, in some sense,
most of these available models provide a fairly good description,
instead of an explanation, to the phenomena.

Worthy to note that, our intention of this work is only to
illustrate whether the statistical effect is important to nucleon
structure, not how well it matches experimental results. So we do
not make any effort to fit the experimental data intentionally.
There is no arbitrary parameter put by hand in our model, and all
parameters are basic statistical quantities. Some of other
statistical models can fit the experimental data better by
introducing many free parameters, however, it weakens the stringency
at a cost.

\section{Statistical approach}

We assume that the nucleon is a thermal system in equilibrium, made
up of free partons. Quarks and anti-quarks satisfy the Fermi-Dirac
distribution, while the gluons obey the Bose-Einstein distribution,
\begin{equation}
f(k^0) = \frac{g_f V}{(2\pi)^3} \frac{1}{e^{\frac{k^0 - \mu_f}{T}}
\pm 1}\;,
\end{equation}
with the upper sign for Fermion, and nether sign for Boson; $g_f$ is
the degree of color-spin degeneracy, hence $g_f = 6$ for quark and
$g_f = 16$ for gluon; $\mu_f$ is its chemical potential, hence
$\mu_{\bar{q}} = - \mu_q$ and $\mu_g = 0$.

We introduce the light-front 4-momentum of the parton $k = (k^+,
k^-, {\bf k}_\perp)$, where $k^+ = k^0 + k^3$, $k^- = k^0 - k^3$,
${\bf k}_\perp = (k^1, k^2)$, and $k^+ = P^+ x = Mx$, where $x$ is
the light-front momentum fraction of the nucleon carried by the
parton. On the trivial assumption that ${\bf k}_\perp$ is
transversely isotropic, we can get PDFs analytically~\cite{lab29}
\begin{equation} \label{fx}
f(x) = \pm\frac{g_fMTV}{8\pi^2}
\left\{\left(Mx+\frac{m_f^2}{Mx}\right)\, \ln\left[1\pm
e^{-\frac{\frac{1}{2}\left(Mx+\frac{m_f^2}{Mx}\right)-\mu_f}{T}}
\right] -2T\text{Li}_2\left(\mp
e^{-\frac{\frac{1}{2}\left(Mx+\frac{m_f^2}{Mx}\right)-\mu_f}{T}}\right)\right\}\theta\left(x-\frac{m_f^2}{M^2}\right)\,
\,,
\end{equation}
\hspace{-0.5cm}where $\text{Li}_2(z)$ is defined as
$\text{Li}_2(z)=\sum_{k=1}^\infty z^k/k^2$, and the step-function
$\theta (x - m_f^2 / M^2)$ originates from the constraint $x \geq
m_f^2 / M^2$~\cite{lab53}.

In practice, the PDFs in a certain system should be constrained with
some conversation laws. For example, in the proton, they are
\begin{equation}
\label{uv} u_V = \int [u(x)-{\bar{u}}(x)]\,\mathrm{d}x=2\;,
\end{equation}
\begin{equation}
\label{dv} d_V = \int [d(x)-{\bar{d}}(x)]\,\mathrm{d}x=1\;,
\end{equation}
\begin{equation}
\label{xnor} \sum_f\int xf(x)\,\mathrm{d}x=1\;.
\end{equation}
For free proton, we also introduce the Gottfried sum,
\begin{equation}
\label{sg} S_G =\int_0^1\frac{F_2^p(x)-F_2^n(x)}{x}\,
\mathrm{d}x=\frac{1}{3}+\frac{2}{3}\int_0^1\left[\overline{u}(x)
-\overline{d}(x)\right]\mathrm{d}x\;,
\end{equation}
whose experimental value is $0.235\pm0.026$~\cite{lab54, lab55}.

Now there are four unknown parameters $T$, $V$, $\mu_u$, $\mu_d$
($M$ is taken as given) and four constraints, i.e.,
Eqs.~(\ref{uv})--(\ref{sg}), thus the parameters can be determined
uniquely. The results for the proton are $T_0=47$~MeV,
$r_0=(3V_0/4\pi)^{1/3}=2.8$~fm, $\mu_u \approx 64$~MeV, and $\mu_d
\approx 36$~MeV. However, the radius $r_0$ seems a little larger
than the realistic value, possibly due to the oversimplified
assumption of the uniform distribution of partons and negligence of
surface effect.

After the PDFs are addressed, the nucleon structure function
$F_2(x)=2xF_1(x)=x\sum_fe_f^2f(x)$ can be attained directly.
Further, with $p$-$n$ isospin symmetry, i.e., $u^n(x)=d^p(x)$,
$d^n(x)=u^p(x)$, $\bar{u}^n(x)=\bar{d}^p(x)$,
$\bar{d}^n(x)=\bar{u}^p(x)$, $g^n(x)=g^p(x)$, we can obtain the
structure function of the neutron as well. Various results, together
with discussions, for PDFs and structure functions are illustrated
in Ref.~\cite{lab29}.

In nucleon, the valence numbers of heavy flavors are zero, then the
chemical potentials of them all vanish, hence no extra parameter is
introduced after adding heavy flavors. However, we found that the
contributions from them are rather small. Further, we identify that
the $s$, $\bar{s}$ asymmetry in the nucleon~\cite{lab56} does not
originate from the pure statistical effect.

\section{Nuclear EMC effect}

The reasonableness and simpleness of the model encourage us to apply
it to nuclear EMC effect~\cite{lab52}. We mainly assume that a
nucleon under a different nuclear circumstance is equivalent to at a
different temperature, and subsequently along with different $V$,
$\mu_u$, and $\mu_d$. Practically, we release the constraint from
the Gottfried sum, i.e., Eq.~(\ref{sg}), and remain
Eqs.~(\ref{uv})--(\ref{xnor}) for protons immersing in the nuclear
environment. In other words, we introduce the temperature $T$ as a
parameter versus the atomic number $A$, then fit the theoretical
ratios of structure functions to experimental data in the EMC region
$x\in[0.2,0.7]$.

The fit is rather impressive~\cite{lab52}. The temperature we get is
about 1$\sim$2 MeV lower in bound nucleons than in free ones, and
jointly the volume is bigger about 5\%$\sim$10\%. Our result is
qualitatively consistent with other models, such as the
$Q^2$-rescaling model and the nucleon swelling model. Worthy to note
that, including the strange quark and taking different masses of it
lead to some slightly difference in results, so the $s$ flavor is
considered as a modification. We also give explicitly the
predictions of PDFs of the nucleons inside different nuclei. The
ratios of the PDFs of iron to deuterium are depicted in
Ref.~\cite{lab52}. They are quite different from the predictions of
other models, i.e., the cluster model~\cite{lab39, lab40}, the pion
excess model~\cite{lab41, lab42, lab43}, and the $Q^2$-rescaling
model~\cite{lab46, lab47, lab48}. And to distinguish various models
and look into the immanent cause of the EMC effect, we suggest more
experiments to identify the PDFs in nuclei, especially for
anti-quarks, the strange quark, and the gluon. Dimuon yield in
Drell-Yan process, semi-inclusive hadron productions in DIS, charmed
quarks production in DIS via the photon-gluon fusion mechanism, and
$\Lambda$-$K$ process, are suggested.

\section{Summary}

We preform a new statistical approach and obtain analytic
expressions of the parton distribution functions (PDFs) in terms of
light-front kinematic variables in the whole $x$ region. There is no
arbitrary parameter or extra corrected term put by hand in our
model, which guarantees the stringency of our conclusion. And then,
we treat the nucleon temperature $T$ as a parameter of the atomic
number $A$ to mimic the nuclear EMC effect, and find that the
nuclear effect can be explained as a shift of $T$; the larger $A$,
the more significant influence. Further, we present the predictions
of PDF ratios for iron as an example. These predictions are rather
different from those of other available models. Experiments are
expected to provide more information of the PDFs in nuclei,
especially for anti-quarks, the strange quark, and the gluon, then
we can test various models better.

All of these show that although the statistical effect is not
everything, it is very important to some aspects of the nucleon
structure and nuclear EMC effect.

\begin{acknowledgments}
This work is supported by National Natural Science Foundation of
China (10721063, 10975003), Hui-Chun Chin and Tsung-Dao Lee Chinese
Undergraduate Research Endowment (Chun-Tsung Endowment) at Peking
University, and National Fund for Fostering Talents of Basic Science
(J0630311, J0730316).
\end{acknowledgments}

\end{document}